\documentstyle{l-aa}
\input{epsf}
\begin{document}

% *************************************************************************
%                                 TITLE
% *************************************************************************
\title{ Electron  $\nu \bar{\nu}$ bremsstrahlung in a
          liquid phase of neutron star crusts}
\author{ P.~Haensel$^1$,
         A.D.~Kaminker$^2$ and D.G.~Yakovlev$^2$}
\offprints{P.~Haensel}
\institute{N. Copernicus Astronomical Center, 
        Polish Academy of Sciences, Bartycka 18,
        00-716 Warszawa, Poland
        \and
         A.F.~Ioffe Institute of Physics and Technology,
         194021 St.Petersburg, Russia}
\thesaurus{02.04.1 - 08.09.3 - 08.14.1}
\date{Received .............; accepted .............}
\maketitle
\markboth{P. Haensel et al: Neutrino pair bremsstrahlung}{}
\label{sampout}

% *************************************************************************
%                             ABBREVIATIONS
% *************************************************************************
\def\la{\;
\raise0.3ex\hbox{$<$\kern-0.75em\raise-1.1ex\hbox{$\sim$}}\; }
\def\ga{\;
\raise0.3ex\hbox{$>$\kern-0.75em\raise-1.1ex\hbox{$\sim$}}\; }
\newcommand{\om}{\mbox{$\omega$}}              % \omega
\newcommand{\th}{\mbox{$\vartheta$}}           % \vartheta
\newcommand{\ph}{\mbox{$\varphi$}}             % \varphi
\newcommand{\ep}{\mbox{$\varepsilon$}}         % \varepsilon
\newcommand{\ka}{\mbox{$\kappa$}}              % \kappa
\newcommand{\dd}{\mbox{d}}                     % d - differential
\newcommand{\vp}{\mbox{\bf p}}                  % vector p (momentum)
\newcommand{\vk}{\mbox{\bf k}}                  % vector k (momentum)
\newcommand{\vq}{\mbox{\bf q}}                  % vector q (momentum)
\newcommand{\vv}{\mbox{\bf v}}                  % vector v (velocity)
\newcommand{\kB}{\mbox{\boldmath $k_{\rm B}$}}  % Boltzmann constant
\newcommand{\vect}[1]{\mbox{\bf #1}}            % vector function

% *************************************************************************
%                               TEXT BODY
% *************************************************************************
\begin{abstract}
Neutrino emissivity from the electron $\nu\bar\nu$
bremsstrahlung in the liquid layers of the neutron star crusts
is studied. 
Nuclear composition of matter in neutron star crusts
is considered for various scenarios of neutron star
evolution. For the deep layers of the crust, the compositions
of cold catalyzed matter, accreted matter and
hot  matter ($T \ga 5 \times 10^9$~K) are
shown to be very different,
and this implies differences in the neutrino 
emissivity at given density and temperature. 
Neutrino-pair bremsstrahlung, due to collisions
of relativistic degenerate
electrons in a Coulomb liquid of atomic nuclei, is considered.
The neutrino energy loss rate
is expressed in a simple form
through a Coulomb logarithm $L$ --
a slowly varying function of density, temperature, and
nuclear composition. Non-Born corrections,
thermal width of electron Fermi sphere, and finite sizes of nuclei
are taken into account.
Implications for
cooling of neutron stars are discussed.
\end{abstract}

%Section 1 ******************************************************
\section{Introduction}
A neutron star crust extends from the stellar surface to
the internal core, up to densities of about $10^{14}$ g~cm$^{-3}$.
The properties of crustal matter are most important
for studying neutron star physics and observational
data (surface thermal X-ray emission, X-ray burst
phenomena, evolution of magnetic fields, etc.).
This article has two goals. First, we
analyze nuclear composition
of matter in neutron star crusts.
Second, we reconsider the
neutrino-pair bremsstrahlung (NPB) in liquid phase of the matter
due to scattering of electrons
on atomic nuclei,
\begin{equation}
        e + (Z,A) \rightarrow e + (Z,A) + \nu + \bar{\nu}.
\label{eq:Reaction}
\end{equation}
NPB is known to be one of the most powerful
mechanisms of neutrino emission
in matter with density lower than the nuclear density,
$\rho_0 = 2.7 \times 10^{14}$ g cm$^{-3}$.

NPB was considered in a number of works (see
Itoh et al. 1989, and references therein).
Let us mention detailed studies of
Festa \& Ruderman (1969),
Flowers (1973), Dicus et al. (1976), 
Soyeur \& Brown (1979), Itoh \& Kohyama (1983), and
Itoh et al. (1984a)
who analyzed NPB due to electron scattering
in liquid and crystalline
dense matter.  Pethick \& Thorsson (1994) have shown that the
band structure effects of electrons in solid matter
drastically suppress the static lattice contribution to NPB.
Finally, the reconsideration of the neutrino energy loss        % ???
rate due to the electron-phonon scattering in the crystal phase
was recently performed by two of us 
(see Yakovlev \& Kaminker 1996).                              % ???

In the present paper, we will reconsider the problem of the
calculation of the neutrino losses from NPB in the hot neutron
star crusts. We will restrict ourselves to the case of a liquid
phase of the crust (temperature above melting temperature at
given density). The actual nuclear composition of the crust
depends on the formation scenario. In contrast to previous
studies, our calculations will be performed for realistic models
of nuclear composition, which are based on specific assumptions
about the formation of the neutron star envelope. 

The paper is organized as follows. In Sect.~2 we
analyze physical conditions and nuclear composition
of neutron star crusts. We discuss those 
differences between various neutron star crusts, which are
important for the NPB emissivities. In Sect.~3 we recalculate
the NPB energy generation rate for
degenerate relativistic electrons and Coulomb liquid of atomic
nuclei. Contrary to the previous works,
we include the non-Born corrections and the effects
associated with thermal width of the electron Fermi sphere.
We express our results in a simple form -- through a
Coulomb logarithm, 
which is a slowly varying function of density, temperature, 
and nuclear composition of matter. In Sect.~4 we discuss NPB
for various models of matter in neutron star crusts,
and indicate possible implications of our results.
General formalism of NPB is presented in Appendix~A,
mathematical aspects of the thermal broadening of
electron Fermi sphere are considered in Appendix~B.
Dynamic screening of the
electron-nucleus interaction
is outlined in Appendix~C.

%
%Section 2 ***************************************************
\section{Composition of neutron star crusts}
%
%Sec. 2.1
\subsection{Physical conditions}
Consider dense matter of a neutron star crust
in a density range
$10^6$ g~cm$^{-3} \ll \rho \la 10^{14}$ g~cm$^{-3}$.
If $\rho <\rho_{\rm ND}\simeq
(4 - 6)\times 10^{11}~{\rm g~cm^{-3}}$, 
matter is composed of bare atomic nuclei
(complete pressure ionization)
immersed in an almost ideal gas 
of  relativistic degenerate electrons. 
This phase of matter constitutes the {\it outer crust} of a neutron
star. Above the neutron drip density, $\rho_{\rm ND}$, some
fraction of neutrons is not bound in nuclei,
and matter is composed
of nuclei, immersed in the electron and neutron gases. The
actual value of $\rho_{\rm ND}$ is weakly model dependent. It
depends also on the scenario of formation of the neutron star
crust. At $\rho \ga 10^{14}~{\rm g~cm^{-3}}$, the
topology of the nucleon distribution can be 
very different from the standard one 
(e.g., neutron gas
bubbles in nuclear matter with a large neutron excess, see
Lorenz et al. 1993 and references therein).
Finally, for $\rho >\rho_{\rm h}\simeq 1.5\times 10^{14}~{\rm
g~cm^{-3}}$, nucleons form a single,
homogeneous phase. Note that the quoted value of
$\rho_{\rm h}$ obtained recently by Lorenz et al.
(1993) is significantly lower than the standard one, based on
earlier calculations (Shapiro \& Teukolsky 1983).
Matter with $\rho_{\rm ND} < \rho < \rho_{\rm h}$
forms an {\it inner crust}.

The state of degenerate electrons in a neutron star crust
is described by the electron
Fermi-momentum $p_{\rm F}$ or relativistic parameter $x$
\begin{equation}
       p_{\rm F} = \hbar (3 \pi^2 n_e)^{1/3}, \hspace{5mm}
       x = \frac{p_{\rm F}}{mc} \approx 1.009
       \left( \frac{\rho_6}{\mu_e} \right)^{1/3},
\label{eq:pF}
\end{equation}
where $\mu_e $ is the number of baryons per electron,
and $\rho_6$ is density in units of $10^6$~g~cm$^{-3}$.
The electron degeneracy (Fermi) temperature is
\begin{equation}
       T_{\rm F} = T_0 \, (\sqrt{1+x^2}-1), \; \;
             T_0 = \frac{mc^2}{k_{\rm B}}
             \approx 5.930 \times 10^9 \;{\rm K}.
\label{eq:TF}
\end{equation}
We will consider strongly degenerate electrons, $T \ll T_{\rm F}$.

The state of the nuclei (ions) is defined by the Coulomb coupling
parameter
\begin{equation}
       \Gamma = \frac{Z^2 e^2}{a k_{\rm B} T} \approx
             0.2254 \frac{Z^{5/3}}{T_8}x,
\label{eq:Gamma}
\end{equation}
where $Ze$ is the nucleus charge,
$a=[3/(4 \pi n_i)]^{1/3}$ is the ion-sphere radius,
and $T_8$ is temperature in $10^8$ K.
At high temperatures, when $\Gamma \ll 1$, the nuclei
constitute a Boltzmann gas. At lower $T$ (higher $\Gamma$)
the gas gradually (without any phase transition)
transforms into a Coulomb liquid. The liquid
solidifies (forms a Coulomb crystal) at $T=T_m$.
For a classical one-component plasma of nuclei,
$T_{\rm m} \, \approx 1.32 \times 10^5 Z^{5/3} (\rho_6 /\mu_e)^{1/3}$~K
corresponds to $\Gamma = 172$ (Nagara et al. 1987).
For high densities and light nuclei,
the crystallization may be affected by
zero-order quantum vibrations of nuclei
(Mochkovitch \& Hansen 1979, Chabrier 1993).

% Sect. 2.2
\subsection{Models of matter in a crust}
If $T\la 5 \times 10^9~{\rm K}\simeq T_0$, one can approximate the
composition of the crust by that calculated at $T=0$. Finite
temperature corrections at $T\sim 10^9$~K result in the presence
of a very small fraction of free neutrons even at
$\rho<\rho_{\rm ND}$, with a negligible effect on the
bulk composition of matter.

The compositions of the outer and inner
crusts depend on the scenario of formation of these outer layers.
We consider three models
of matter in a neutron  star crust.
The first model is based on the
assumption that matter is in its ground state ({\it cold catalyzed
matter}). The second one is the model of {\it accreted matter}
valid when the neutron star accreted a sufficient
amount of matter during its life.
Both models are appropriate for $T \la 5 \times 10^9$~K.
In both cases the nuclei
are assumed to constitute a one-component plasma ($A$, $Z$),
which behaves either as a Coulomb liquid ($T>T_{\rm m}$) or as a Coulomb
body-centered cubic crystal ($T<T_{\rm m}$).
In addition we consider (Sect. 2.4) the model of {\it hot matter}
valid for $T \ga 5 \times 10^9$~K.

Some uncertainties in
calculations of nuclear composition
come from extrapolation of the
laboratory nuclear physics to very large neutron
excesses and huge pressures in neutron star crusts.
These uncertainties are more important
for the inner crust, and they increase with increasing $\rho$.

% Sec. 2.3
\subsection{Cold catalyzed and accreted matter,
            $T \la 5 \times 10^9$~K}
Let us first consider the models of cold catalyzed and accreted matter
in the {\it  outer crust} ($\rho<\rho_{\rm ND}$).

At finite temperature the first model corresponds to a complete
thermodynamic equilibrium. This assumption is certainly valid at
$T \ga 10^{10}$~K (neutron star birth); its validity is extrapolated
to the later stages of neutron star cooling.
This determines the composition at a given pressure (density).
In our analysis of the outer crust, we use
the model of Haensel \& Pichon (1994) based on the most
recent experimental data on neutron rich nuclei.

The model of accreted matter represents
the case, in which matter {\it is not}
in the ground state. At typical accretion
rates onto a neutron star in a close binary system, freshly
accreted hydrogen-rich matter burns into helium. The latter, in turn,
burns explosively into $^{56}$Ni, which transforms eventually
into $^{56}$Fe. The products
of thermonuclear burning are subsequently compressed under the
weight of accreted matter at relatively low temperature.
For instance, at the
accretion rate $\sim 10^{-10}~M_\odot/{\rm yr}$ typical temperature
within the crust is a few times of $10^8$~K.
Accordingly the only further nuclear processes in the steadily
formed `new  outer crust' are electron captures, which eventually
lead to neutron drip at $\rho_{\rm ND} \simeq  6 \times 10 ^{11}
$ g~cm$^{-3}$ (e.g., Haensel \& Zdunik 1990).
Note, that at the accretion rate
$10^{-10}~M_\odot/{\rm yr}$ it takes $10^5$~yr to replace the
original outer crust, built of catalyzed (ground state)
matter, by the new one.

The compositions of the accreted and cold catalyzed
outer crusts are vastly different.
At $\rho \simeq 10^{11}$~g~cm$^{-3}$
the values of $Z$ and $A$ in the accreted matter
are less than half of those in
the ground state.
For instance, at the neutron drip
we have $Z$=18, $A$=56 for the accreted matter
(Haensel \& Zdunik 1990) and $Z$=35, $A$=118 for the cold
catalyzed  matter (Haensel \& Pichon 1994).
The neutron drip density, which separates
the {\it outer} and {\it inner} crusts,
is rather insensitive to the
scenario of the outer crust formation (cold catalyzed matter
or accreted matter) and/or to the model of nuclei present in the
matter. Irrespectively of the model, one gets
$\rho_{\rm ND} = (4 - 6) \times 10^{11}$~g~cm$^{-3}$
(Shapiro \& Teukolsky 1983, Haensel et al. 1989,
Haensel \& Zdunik 1990).

Now consider  the models of
ground state and accreted matter
in the {\it inner crust} ($\rho > \rho_{\rm ND}$).
In both cases we assume that (for
$\rho \la 10^{14}$~g~cm$^{-3}$) matter is composed of
spherical, neutron rich
nuclei $(A, Z)$ immersed in the electron and neutron gases. 
Because of relatively low temperature, the only processes 
involved in the formation of the ``inner accreted crust'' are 
electron captures, neutron emission and absorption, 
and pycnonuclear  fusion. 

For the ground state matter at $\rho > \rho_{\rm ND}$,
we use the results of Negele \& Vautherin (1973).
The model of accreted crust is taken from Haensel \& Zdunik (1990).
We have constructed a specific model of an accreted crust for
a neutron star which accreted $6\times 10^{-4}~M_\odot$.
The bottom of the accreted layer is found at
$\rho\simeq 1.1\times 10^{13}$~g~cm$^{-3}$.

Both models are developed for a one component
plasma (single $A,Z$ nuclide at a given pressure). This
implies the onion structure of the crust, with fixed values
of $A,Z$ within certain pressure intervals.

Figures 1  and 2 present the relevant parameters
of the ground state and accreted crusts
versus density. The
discontinuities in the values of $(A,Z)$ are due to the shell
and pairing effects in the binding energies of nuclei.
Notice a strong dependence of $Z^2/A$ (which
is important for NPB, Sects.~3 and 4) on the
crust formation scenario for $10^9~{\rm g \; cm}^{-3} \la
\rho \la  10^{13}~{\rm g~cm^{-3}} $. The values of $Z^2/A$
for the accreted crust are about $2-3$ times
lower than for the ground state.
The differences of the melting temperature
are even higher (a factor of $\sim 6$ at
$\rho \sim 10^{12}~$g~cm$^{-3}$, Fig.~2).
The melting temperature of the ground
state matter for $\rho>10^{12}$~g~cm$^{-3}$ is
$\ga 5 \times 10^9$~K. At such temperatures the
thermal corrections to the composition of matter become important.

%%%%%%%%%%%%%%%%%%%%%%%%%%%%%%%%%%%%%%%%%%%%%%%%%%%%%%%%%%%%%%%%%%%%
Let us notice that in the case of accreted crust at
$\rho>10^{10}~{\rm g~cm^{-3}}$, the ion plasma temperature,
$
T_{\rm p}=\hbar \omega_{\rm p}/k_{\rm B}=
(\hbar/k_{\rm B})
(4\pi Z^2 e^2 n_{\rm i}/m_{\rm i})^{1/2}~$,
 where $n_{\rm i}$ is the number density of nuclei (ions) and 
 $m_{\rm i}$ is their
mass, becomes larger than the melting temperature of {\it
classical} plasma. This seems to indicate, that  
corrections to $T_{\rm m}$, resulting from quantum (zero--point)
vibrations  of 
 nuclei, might become significant. However, a quantitative
analysis, based on the formulae derived by 
Chabrier (1993) (who corrected formulae 
 obtained by Mochkovitch \& Hansen
(1979) ) shows, that even in the most extreme cases, corresponding
to minima of $T_{\rm m}$ at $\rho \ga 10^{12}~{\rm g~cm^{-3}}$
(see Fig. 2), quantum
corrections to $T_{\rm m}$ were about a few percent. 
 In the case of the ground state of matter we have
always $T_{\rm p}<T_{\rm m}$, and so the quantum corrections to
$T_{\rm m}$ are even smaller. 
%%%%%%%%%%%%%%%%%%%%%%%%%%%%%%%%%%%%%%%%%%%%%%%%%%%%%%%%%%%%%%%%%%%%%%%%

%******************************************************************
%                                                       FIGURE 1.
\begin{figure}
\begin{center}
\leavevmode
\epsfxsize=8.0cm \epsfbox{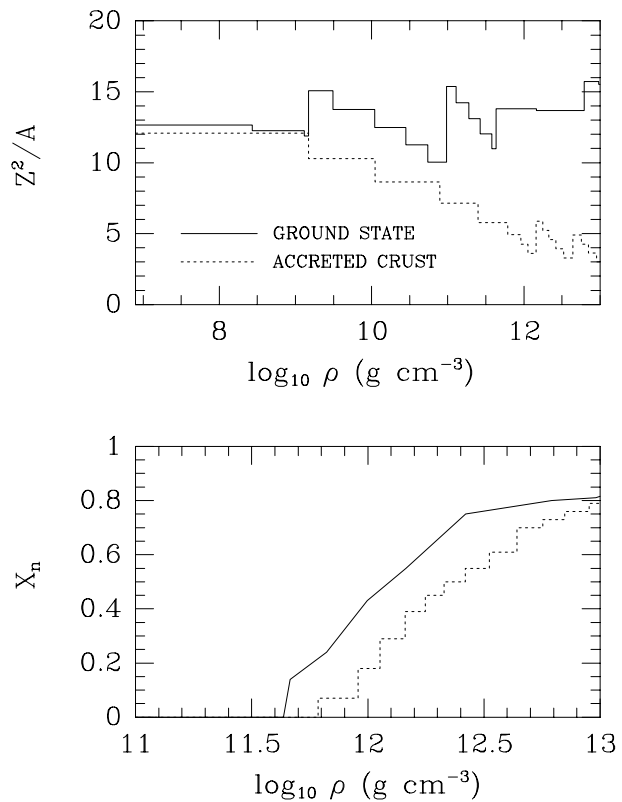}
\end{center}
\caption[ ]{
Nuclear factor $Z^2/A$, important for NPB, and 
the mass fraction of free
neutrons, $X_{\rm n}$ (related to mass fraction  of 
atomic nuclei, $X_A$,  by $X_A=1-X_{\rm n}$),
 versus density $\rho$ (in g~cm$^{-3}$), 
 for two models of neutron star crust. Solid line: ground state
of matter; dotted line: accreted crust.
}
\label{fig1}
\end{figure}
%******************************************************************

%******************************************************************
%                  FIGURE 2.
\begin{figure}
\begin{center}
\leavevmode
\epsfxsize=8.0cm \epsfbox{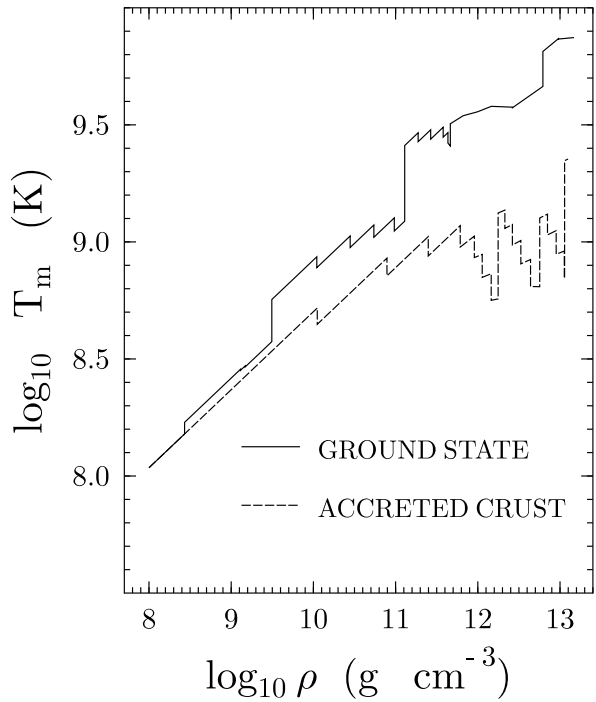}
\end{center}
\caption[ ]{
Melting temperature $T_m$ (K), versus density (in g~cm$^{-3}$), for the
same models as in Fig.~1.
}
\label{fig2}
\end{figure}
%******************************************************************
%\vfill
%\end{document}
% Sec. 2.4
\subsection{Hot matter, $T \ga 5 \times 10^9$~K}
For $T> 5 \times 10^9~{\rm K}$,
the thermal effects strongly modify
the nuclear composition. Both shell and
pairing effects are washed out from the abundances of nuclei.
Also, reshuffling of nucleons is no longer blocked efficiently
by Coulomb barriers, and one can assume complete thermal
equilibrium. We treat hot matter as a mixture of neutrons,
protons, electrons, positrons and nuclei. In addition, we represent
light nuclei by the $\alpha$ particles, and heavy nuclei by a
single species $(A,Z)$ (e.g., Burrows \& Lattimer 1984).

We describe the nuclei in hot dense matter, using the
model of Lattimer \& Swesty (1991), with a specific choice of
the equilibrium incompressibility of nuclear matter, $K_{\rm s}
=180~$MeV. We assume nuclear equilibrium, as well as beta
equilibrium of matter. The assumption of nuclear
equilibrium is justified by high temperature. Beta equilibrium
is adopted for simplicity: a very rapid
cooling of matter at highest temperatures can produce some
deviations from beta equilibrium.

Figure 3 shows the composition of matter for $T$=
$5 \times 10^9$~K , $8 \times 10^9$~K, and $1.2 \times 10^{10}$~K.
One can see a temperature dependence
for $\rho \la 10^{10}$~g~cm$^{-3}$.
On the other hand, for
$\rho \ga 10^{12}$~g~cm$^{-3}$, the temperature effects are negligible.
If $T \sim 5 \times 10^9$~K,
the thermal effects are weak and produce a small
admixture of free neutrons at $\rho<\rho_{\rm ND}$. The
fraction of free neutrons for $\rho<\rho_{\rm ND}$ rapidly
increases with growing temperature, and small fractions of
$\alpha$-particles and free protons
appear at $\rho< 10^{10}$~g~cm$^{-3}$
($T=8\times 10^9$~K). If $T=1.2 \times 10^{10}$~K,
the nuclei evaporate completely for
$\rho \la 10^9$~g~cm$^{-3}$, and the mass fraction contained in free
neutrons is significant at any density under discussion. 
 For $\rho \ga 10^{12}$~g~cm$^{-3}$, the temperature effects are negligible.

In contrast
to the nuclear composition, the value of $Z^2/A$
depends on temperature rather weakly (Fig.~4).
Its density dependence is smooth, because nuclear pairing and shell
effects are washed out by temperature.

%******************************************************************
%                                                       FIGURE 3.
\begin{figure}
\begin{center}
\leavevmode
\epsfxsize=8.0cm \epsfbox{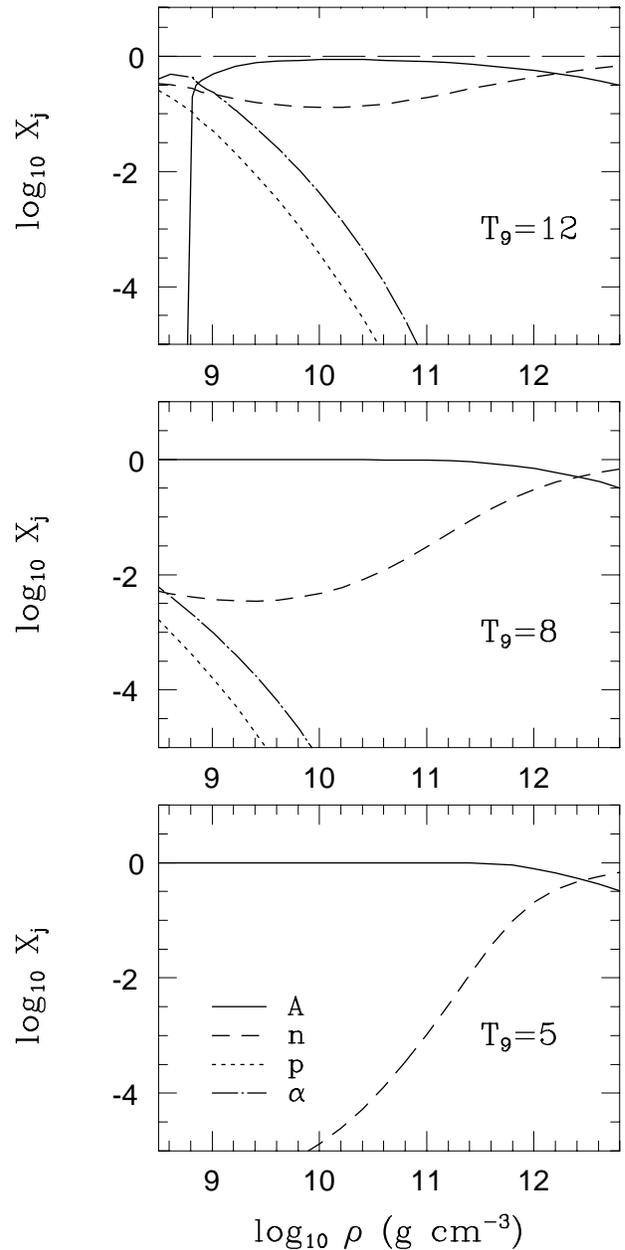}
\end{center}
\caption[ ]{
Mass fractions of various particles versus density at
different temperatures $T_9=T/(10^9~{\rm K})$ for
the model of hot matter, described in the text.
}
\label{fig3}
\end{figure}
%******************************************************************

%******************************************************************
%                                                       FIGURE 4.
\begin{figure}
\begin{center}
\leavevmode
\epsfxsize=8.0cm \epsfbox{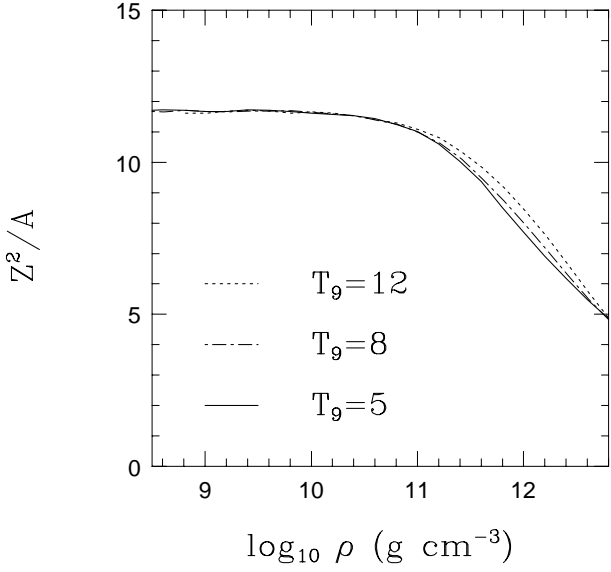}
\end{center}
\caption[ ]{
The nuclear factor $Z^2/A$, important for NPB, versus 
matter density $\rho$, for different
temperatures of hot matter.
}
\label{fig4}
\end{figure}
%******************************************************************

% Section 3 ********************************************************
\section{NPB energy loss rate}
%
% sec. 3.1 -----------------------------------------------------
\subsection{General remarks}
Let us consider NPB (Eq.\,(\ref{eq:Reaction})) 
in the case of
relativistic degenerate electrons
in liquid phase of matter of a neutron star
crust ($T>T_{\rm m}$). NPB has been studied by many authors
(Sect.~1). Thus we omit the details
and outline the derivation of the neutrino energy generation rate
in Appendix~A.

At the first stage, we adopt the Born approximation and neglect
energy exchange between electrons and atomic nuclei. We will discuss
the non-Born corrections in Sect.~3.3,
and the dynamic effects of nucleus response in Appendix~C.
We will mostly use the units in which $c=\hbar=k_{\rm B}=1$.

Let $P=(\ep, \vp)$ be the electron 4-momentum
in the initial state, $P'=(\ep',\vp')$ the electron
4-momentum in the final state, and $Q=(\Omega,\vq)$ be the 4-vector
of momentum transfer to a nucleus.
Since the energy transfer to a nucleus
is neglected we assume $\Omega$=0
throughout Sects.~3 and 4 (although we present general
equations in Appendices~A and C).
Let $K_1=(\om_1,\vk_1)$ and $K_2=(\om_2,\vk_2)$ be
the 4-momenta of neutrino and antineutrino, respectively,
and $K=K_1+K_2=(\om,\vk)$ be the 4-momentum
of the neutrino pair ($\om=\om_1+\om_2$ and $\vk=
\vk_1+\vk_2$). Energy-momentum conservation implies
\begin{equation}
        P=P'+Q+ K.
\label{eq:MomentumConservation}
\end{equation}

As shown in Appendix~A,
the neutrino energy production rate $Q_{\rm Brem}$
(ergs~cm$^{-3}$~s$^{-1}$) in a relativistic electron gas is
\begin{eqnarray}
          Q_{\rm Brem} & = & \frac{G_F^2 C_+^2\,n_i }{12 (2 \pi)^{10}}
             \int \! \dd \vk \int \! \dd \vp
             \int \! \dd \vp' \, |U(\vq)|^2
                           \nonumber \\
                     & \times  &
                  f (1-f') \frac{\om}{ \ep \ep'} J_+ \, ,
          \label{eq:Qgeneral}
\end{eqnarray}
where $G_{\rm F}=1.436 \times 10^{-49}$ ergs~cm$^3$ 
is the Fermi weak coupling constant,
$C_+^2=C_V^2+C_A^2+2({C'}_V^2+{C'}_A^2)$
is expressed in terms of vector and axial vector
constants (Appendix~A) and takes into account generation of
electron neutrinos ($C_V$ and $C_A$;
weak neutral + charged currents)
and also of muonic and tauonic neutrinos (${C'}_V$ and ${C'}_A$;
neutral currents),
$U(\vq)$ is the Fourier transform
of the electron-nucleus Coulomb potential,
$J_+$ is the spin averaged squared matrix element
in the limit of ultrarelativistic electrons. Furthermore,
\begin{equation}
      f= \left[1+\exp \left( {\ep - \mu \over T} \right) \right]^{-1}
\label{eq:FermiDirac}
\end{equation}
is the Fermi-Dirac function for the initial electron, 
$f'\equiv f(\ep')$ is the same function for the final electron, and
$\mu$ is the electron chemical potential.
Integration in (\ref{eq:Qgeneral}) is to be
carried out over the domain where $K^2 \geq 0$.

It is convenient to express $Q_{\rm Brem}$ in the form
(in the ordinary physical units)
\begin{eqnarray}
      Q_{\rm Brem} & = & \frac{8 \pi G_{\rm F}^2 Z^2 e^4 C_+^2}
          {567 \hbar^9 c^8} (k_{\rm B} T)^6 n_i L
                 \nonumber \\
               &  \approx  & 3.229
         \times 10^{17}\, \rho_{12} \,X_A\,
         \frac{Z^2}{A} \,
         T_9^6 L \; {\rm ergs \; s^{-1} \; cm^{-3}},
\label{eq:QthroughL}
\end{eqnarray}
where $X_A$ is the mass
fraction contained in nuclei, and $\rho_{12}$ is density in
the units of $10^{12}$~g~cm$^{-3}$.
Numerical expression for $Q_{\rm Brem}$ is obtained for
the emission of $\nu_e$, $\nu_\mu$, $\nu_\tau$,
$C_+^2 \approx 1.675$, with the Weinberg angle
$\sin^2 \Theta_W \approx 0.23$ (see Appendix~A).
In Eq.~(\ref{eq:QthroughL})
we have introduced the dimensionless quantity $L$.
We will see (Sects.~3.3 and 3.4)
that $L \sim 1$ is a slowly varying function of density,
temperature and nucleus parameters which has
meaning of a {\it Coulomb logarithm}.
Thus the problem reduces to evaluating $L$.
Note that if we replace
$3.23 \times 10^{17}\,L$  by $2.18 \times 10^{17}$,
then Eq.~(\ref{eq:QthroughL}) transforms into
the well known formula of Soyeur \& Brown (1979).

%
%**********************************************************
%sec. 3.2 ------------------------------------------------------
\subsection{Screened Coulomb potential}
The squared Fourier transform of the
Coulomb potential
screened by the plasma polarization
(which enters Eq.~(\ref{eq:Qgeneral})) can be written as
\begin{equation}
         |U(\vq)|^2  =  { (4 \pi Z e^2)^2 \over
                    q^4 | \epsilon(q) |^2 } \, S(q) |F(q)|^2 .
\label{eq:Uq}
\end{equation}
In this case $F(q)$ is a nuclear formfactor
which takes into account the proton charge 
distribution in a nucleus, $S(q)$ is the static structure
factor of ions that describes the
ion screening due to ion-ion correlations, and
$\epsilon(q)$ is the static longitudinal dielectric function
of the electron gas, which accounts for the electron screening.

It is easy to show that the electron screening is always
static for the conditions of study.
The dielectric function of
the degenerate electrons was derived by Jancovici (1962).
In the ultrarelativistic limit
\begin{eqnarray}
       \epsilon(q) &  =  &  1 + \frac{k_{\rm TF}^2}{q^2}
           \left( \frac{2}{3} + \frac{1-3y^2}{6y}
           \ln \left| \frac{1+y}{1-y} \right|  \right.
                   \nonumber   \\
                  &  +   &    \left.
           \frac{y^2}{3} \ln \left| \frac{y^2}{1-y^2}
           \right| \right),
\label{eq:epsilon}
\end{eqnarray}
where
\begin{equation}
       y=\frac{q}{2p_{\rm F}}, \hspace{5mm}
       k_{\rm TF}=p_{\rm F}
       \left( \frac{ 4 e^2}{\pi \hbar v_{\rm F}} \right)^{1/2},
\label{eq:kTF}
\end{equation}
$k_{\rm TF}$ being the electron static
screening momentum, and $v_{\rm F}$ the
electron Fermi velocity. In the long-wavelength
limit ($q \ll p_{\rm F}$),
one has
\begin{equation}
      \epsilon(q) = 1 + \frac{k_{\rm TF}^2}{q^2},
\label{eq:epsilonTF}
\end{equation}
which corresponds to the Debye screening.
Actually this approximation appears
to be quite satisfactory
although we have used exact
dielectric function (\ref{eq:epsilon}) in numerical calculations.

For a spherical nucleus with a uniform proton core, 
\begin{equation}
       F(q) = \frac{3}{(q r_0)^3}
               \left[ \sin(q r_0) - q r_0 \cos(q r_0) \right].
\label{eq:Formfactor}
\end{equation}
Here $r_0$ is the core radius
which can be smaller than the nucleus radius, for
neutron-rich nuclei. One has $F(q)=1$, for point-like nuclei.

The structure factor $S(q)$ in (\ref{eq:Uq})
has been calculated by many authors (e.g., Itoh et al.
1983) for a one-component classical plasma
of ions with the uniform electron background.
In a strongly coupled plasma
($\Gamma \ga 1$), $S(q)$
cannot be evaluated analytically.
Its general feature is that it is greatly suppressed
in the long-wavelength limit, $qa \la 1$ ($a$ is
defined in Eq.~(\ref{eq:Gamma})). Thus
$S(q)$ produces the screening of the
Coulomb interaction with the screening momentum $ \sim 1/a$.

%%%%%%%%%%%%%%%
 Note that actually the structure factor can be influenced
by response of  neutron gas, which surrounds atomic nuclei
in dense matter, to the motion of nuclei.
This effect has not been considered so far in the literature.
 We will use
the conventional structure factors (Itoh et al. 1983),  assuming
that the inclusion of this  effect can be reduced to 
proper choice  of the fraction  of free neutrons
and the number of neutrons bound in nuclei (i.e., 
proper determination of
the nucleus mass number).
%%%%%%%%%%%%%%%

The electron and ion screenings suppress the Coulomb
interaction for small $q$
while the nuclear formfactor reduces the
interaction for large $q \, \ga 1/r_0$. For typical parameters
of dense stellar matter, the ion screening is more
efficient than the electron one ($k_{\rm TF}\, < \, a^{-1}$).

%
%sec. 3.3 ----------------------------------------------------
\subsection{Coulomb logarithm}
Using Eqs. (\ref{eq:Qgeneral}, 
\ref{eq:QthroughL}) and (\ref{eq:Uq})
we come to the general expression for the Coulomb logarithm
(here and below again $\hbar = c = k_{\rm B} = 1$):
\begin{eqnarray}
     L  & = & \frac{189}{2^{11} \pi^9 T^6}
        \int \dd \vk \; \dd \vp \; \dd \vp'
        \frac{1}{q^4 \ep \ep'} \;
        {S(q) |F(q)|^2 \over |\epsilon(q)|^2}
              \nonumber \\
                & \times    &
              f(1-f') \; \om J_+.
\label{eq:Lgeneral}
\end{eqnarray}
Let us simplify the integration.
Since the electrons are strongly degenerate, the main contribution
to $L$ comes from those electron transitions
in which the electron momenta $\vp$ and $\vp'$ lie
in the narrow thermal shell around the Fermi surface,
$|\ep - \mu| \la T$  and  $|\ep' - \mu| \la T$.
Let $q_s \sim 1/a$ be the typical screening momentum
(Sect.~3.2). We assume that $q_s \ll p_{\rm F}$ to simplify
our analysis. 

Let $\vq = \vq_t+ \vq_r$, where $\vq_t$     
corresponds to purely elastic Coulomb scattering
while $\vq_r$ takes into account inelasticity.
 Here the inelasticity means that the length of electron momentum
changes slightly due to the Coulomb interaction
although the energy transfer to the nucleus is absent.
The process is kinematically allowed since two
neutrinos are also involved in energy-momentum conservation.
Let us also introduce the vector 
$\vp''= \vp - \vq_t= \vp'+\vq_r + \vk$
which is directed along
$\vq_r$ but has the same length as $\vp$.

Note that $q_t = 2p \sin(\theta/2) \approx 2 p_{\rm F} \sin(\theta/2)$,
where $\theta$ is an angle
between $\vp$ and $\vp''$. From geometrical consideration
we obtain
\begin{equation}
       q^2 = ( \vq_t + \vq_r )^2 = 
           q_t^2 + q_r^2 - q_t^2 
          \left( {q_r \over p_{\rm F}} \right).
\label{eq:q}
\end{equation}

The neutrino-pair momentum $\vk$ can be presented as
$\vk = \vk_t + \vk_r$, where $\vk_r$ and $\vk_t$
are the orthogonal vector components parallel and 
perpendicular to $\vp''$,
respectively. Strong electron degeneracy implies that
$q_r$ and $k$ are much smaller than $p_{\rm F}$
although $q_t$ can be comparable with $p_{\rm F}$
for large-angle electron scattering events.
Therefore, NPB is accompanied by nearly elastic
Coulomb scattering.
Then the neutrino-pair energy is
\begin{equation}
       \omega = { \ep^2 - {\ep'}^2 \over \ep + \ep'} 
       \approx   k_r + q_r.
\label{eq:omega}
\end{equation}

The condition $K^2= \om^2-\vk^2 \approx k_0^2 - k_t^2 >0$
requires  $k_0 \geq k_t $, where $k_0^2=q_r(2 \om-q_r)$, or
$q_r > 0$,  $\om > q_r/2$.
Furthermore, we set $\dd \vp = \dd \Omega \ep^2 \dd \ep$, 
where $\dd \Omega$ is solid angle element in the direction 
of vector $\vp$. 
The integration over $\ep$ under strong electron
degeneracy is standard:
\begin{equation}
       \int \dd \ep f(1-f') = \frac{\om}{{\rm e}^{ \omega / T} -1}.
\label{eq:FermiDiracInt}
\end{equation}
The integration over $\vp'$
can be replaced by the integration over $\vq$, with $\dd \vq =
2 \pi q_t \, \dd q_t \, \dd q_r$.
Note also, that $\dd k_r = \dd \om$.
This yields
\begin{eqnarray}
   L & = & \frac{189}{2^{11} \pi^9 T^6} \; 8 \pi^2            
        \int_0^\infty  \dd q_r \; \int \dd q_t \; q_t \;
          {S(q) |F(q)|^2 \over q^4 | \epsilon(q)|^2} \;
\nonumber       \\
       & \times  &  \int_{q_r/2}^\infty \dd \om \; \frac{\om^2}
        { {\rm e}^{\omega / T} -1}
       \int_0^{k_0}  \dd k_t \, k_t \,
       \int_0^{2 \pi} \dd \varphi   \; J_+,
\label{eq:LthroughJ}
\end{eqnarray}
where $\varphi$ is an azimuthal angle of $\vect{k}$
with the polar axis along $\vp''$.
The expression for $J_+$ is given by
Eq.~(\ref{eq:AJ}) in Appendix~A. Using this expression
and the relation
$\vq_t\vk=-2p_{\rm F} k_r \sin^2(\theta/2) +
p_{\rm F} k_t \sin \theta \cos \varphi $,
we can integrate over $\varphi $ and $k_t$
and present $L$ in the form
\begin{eqnarray}
      L & =  &
          {1 \over T} \int_0^{2p_{\rm F}} \; \dd q_t \; q_t^3  \;
            \int_0^\infty \dd q_r \;
             {S(q) |F(q)|^2 \over  \ q^4 |\epsilon(q)|^2}
                     \nonumber  \\
                     & \times    &
              R_T(y,u) \;  R_{\rm NB}(q_t),
\label{eq:L}
\end{eqnarray}
where $y=q_t/(2 p_{\rm F})= \sin(\theta/2)$, $u=q_r/T$,
and $q$ is given by (\ref{eq:q}). Here we
introduce the factor $R_{\rm NB}$ to take into account
deviations from the Born approximation (see below).
The integration over $q_r$ describes the {\it thermal effects}, i.e.,
a weak inelasticity in Coulomb
scattering due to a finite width of
the thermal shell around the electron Fermi surface.
In Eq.~(\ref{eq:L}) these effects are controlled by the function
\begin{eqnarray}
       R_T(y,u)  & = & \frac{189}{16 \pi^6}
            \left\{  \int_{u/2}^{u/(2y^2)} \dd v
            \frac{ v^2}{ {\rm e}^v -1}
            \left(v - \frac{u}{2} \right)^2  \right.
                 \nonumber \\
                  & \times    &
            \left[1 - {2 \over 3u} \; {y^2 \over 1-y^2}
            \left(v - {u \over 2} \right) \right]
                  \nonumber \\
       &+&   {u \over 2}\, {1-y^2 \over y^2}
            \int_{u/(2y^2)}^\infty \dd v \;
            {v^2 \over {\rm e}^v-1 }
                   \nonumber \\
                   &  \times   &  \left.
            \left[v - {u \over 2} - {u \over 6} \, {1-y^2 \over y^2}
            \right] \right\},
\label{eq:RT}
\end{eqnarray}
where $v= \om /T$. The thermal
effects introduce an additional screening
of the Coulomb interaction 
with the screening momentum $\sim T$.
The properties of $R_T(y,u)$ are analyzed in
Appendix~B, where we present also a convenient fitting expression
for this function.

Finally, let us discuss the non-Born correction
$R_{\rm NB}$ in (\ref{eq:L}).
Since dense stellar matter contains heavy
nuclei (Sect.~2),
the Born approximation used in all
previous works
is not very accurate. Exact calculation of
the NPB rate beyond the Born approximation
is difficult but we propose an 
approximate treatment of the non-Born terms, which is explained
below. 

A process of NPB consists
of two stages: electron scattering
on a nucleus, and neutrino-pair emission. As seen from
the above results the electron scattering
is much stronger (i.e., it is accompanied by larger momentum
transfers) than the neutrino-pair emission except possibly
for the small-angle scattering. However one cannot expect
large deviations from the Born approximation for small-angle
scattering (Berestetskii et al. 1982).
Thus we can use the well known method of soft photons
(Berestetskii et al. 1982) and claim that
the NPB rate is proportional to the product of the
cross section of the elastic electron - nucleus scattering
and the probability of the neutrino-pair
emission. The pair emission involves weak interaction
and cannot be affected strongly by the Born approximation.
Hence the main impact of the Born approximation is on the
elastic scattering cross section. We expect that the
non-Born correction factor is
$R_{\rm NB}(q_t) = \sigma(q_t)/\sigma_{\rm Born}(q_t)$,
where $\sigma(q_t)$ is the exact elastic cross section
with the momentum transfer $q_t$, and
$\sigma_{\rm Born}(q_t)$ is the Born cross section.
The factor $R_{\rm NB}(q_t)$ has been
calculated by many authors (e.g., Doggett \& Spencer, 1956),
and its inclusion into the Coulomb
logarithm is straightforward. 

%sec 3.4 -----------------------------------------------------
\subsection{Low- and moderate-temperature cases}
We have reduced the problem of calculating the Coulomb logarithm
to a two dimensional integration in (\ref{eq:L}).
One can distinguish two cases:
{\it the low-temperature case} when
temperature is lower than the
Coulomb screening momentum,
$T \ll q_s $, and {\it the moderate-temperature case} when
$q_s \la T \ll T_{\rm F}$.

All previous studies of NPB from the relativistic
degenerate electrons
have used the approximations appropriate
to {\it the low-temperature case}.
In this case the main contribution into
$L$ comes from the values of $q_t \gg T$. Then
$R_T$ is a sharp function of $q_r$ which
decreases rapidly with increasing $q_r \ga T$
(Appendix~B). Therefore, we can set $q_r=0$ in the remaining
functions under the integral (\ref{eq:L}),
and the integration over $q_r$ is performed with the aid
of Eq.~(\ref{eq:BIntR}):
\begin{equation}
      L= \int_{0}^{2p_F} q^3 \; \dd q \,
          {S(q)|F(q)|^2 \over q^4 |\epsilon(q)|^2} \;
           R_c(y) \; R_{\rm NB}(q)~,
\label{eq:LlowT}
\end{equation}
where $q=q_t$, $y=q/(2p_{\rm F})$  and
\begin{equation}
          R_c(y) = 1 + \frac{2 y^2}{1- y^2} \ln(y).
\label{eq:Rc}
\end{equation}
Equations (\ref{eq:QthroughL}) and (\ref{eq:LlowT})
with $R_{\rm NB}=1$ reproduce the familiar
NPB rate for degenerate and
relativistic electrons. For example, these equations
can be obtained from
the results of Itoh \& Kohyama (1983),
taking into account that
\begin{equation}
    \int_0^1 \dd x \, \frac{x(1-x^2)}{y \xi} \ln \left|
    \frac{yx + \xi}{yx - \xi} \right| = \frac{2}{3} \, R_c(y) \,
    \frac{1}{1- y^2},
\label{eq:int}
\end{equation}
where $\xi =\sqrt{1- x^2(1-y^2)}$.
The quantity $F_{\rm liquid}$ introduced by Itoh \& Kohyama (1983)
is equal to $(2/3) L$, and their function
$I_2(q)=(8/3) \; y^2 \, R_c(y)$.

We see that $L$ has, indeed, the meaning of {\it the Coulomb
logarithm}. If the screening of the Coulomb potential were
weak ($S(q)=\epsilon(q)=1$) and
$R_c \approx 1$ (which is true for $q \ll p_{\rm F}$),
then $L$ would acquire a familiar
logarithmic divergency at small $q$. The
divergency is eliminated
due to the screening functions (Sect.~3.2).
The function $R_c$, Eq.~(\ref{eq:Rc}),
introduces additional screening which comes from
the squared matrix element of the NPB reaction. We have
$R_c \to 1$ for small-angle scattering ($y \to 0$),
while $R_c \approx (1 - y^2)/2  \to 0$
for backscattering ($y \to 1$).
Therefore $R_c$ suppresses the backscattering (just as $F(q)$)
which is natural for relativistic
electrons (Berestetskii et al. 1982).

The {\it low-temperature} Coulomb logarithm $L$
is expressed as a simple 
one dimensional integral (\ref{eq:LlowT}),
which is easily computed
(Sect.~3.5) for any parameters of dense stellar matter.

In the {\it moderate-temperature}  case ($q_s \la T \ll T_{\rm F}$),
one should deal with
two dimensional integration (\ref{eq:L})
which is also easy once
the thermal function $R_T(y,u)$ is known (Appendix~B).

%sec. 3.5 ------------------------------------------------
\subsection{Numerical results}
In the degenerate electron gas and Coulomb
liquid of atomic nuclei, the Coulomb logarithm
depends actually on four parameters:
on nuclear charge number $Z$ that defines
the ion screening (Sect.~3.2),
on the proton core nuclear radius $r_0$ (see (\ref{eq:Formfactor})),
on density $\rho$
and $T$. The ion coupling parameter
$\Gamma$, Eq. (\ref{eq:Gamma}),
is expressed through $Z$, $\rho$ and $T$.
In the ultra-relativistic limit ($x \gg 1$) the four parameters
can be replaced by three dimensionless parameters
\begin{equation}
    L=L(Z,\eta,t), \;\; \eta= \frac{r_0}{a}, \; \;
    t= \frac{k_{\rm B}T}{2p_{\rm F} c} \approx 
    \frac{T}{2T_{\rm F}},
\label{eq:dimensionless}
\end{equation}
where $a$ is the ion-sphere radius (see
 Eq. (\ref{eq:Gamma})), and $T_{\rm F}$ is given by Eq.~(\ref{eq:TF}).

We have calculated $L$ from Eq. (\ref{eq:L})
for $Z$ = 10, 20, 30, 40, 50, and for wide ranges of
$\eta \la 0.2$ and $t \la 0.1$ in the Coulomb liquid
($T > T_m$, i.e., $t > 1.1 \times 10^{-5} Z^{5/3}$,
Sect.~2.1). The structure factor $S(q)$
has been taken from Itoh et al. (1983).
Typical results are presented in Figs. 5 and 6.

At small $t$, our values of $L$ tend to those obtained
in the low-temperature approximation 
(cf., curves 3 and 4 in Figs. 5 and 6).
At higher $t$, the
deviations from the latter approximation are
quite pronounced. In the low-temperature approximation,
$L$ depends on $T$ only through the structure factor
$S(q)$ (which slightly varies with $\Gamma$),
and $L$ grows slowly with $t$ due to
weakening of the ion screening.
An inclusion of the thermal effects
(Sects.~3.3 and 3.4) leads to a stronger and non-monotonic
temperature dependence of $L$ which can be explained
as follows. The thermal effects are described by the
function $R_T$ which, generally, introduces an
additional `thermal screening' of the Coulomb interaction
and tends to suppress $L$ and the NPB rate. However the thermal effects
act within the thermal shell near the electron
Fermi surface.
Deviations of the electron momenta from the Fermi sphere
can decrease the momentum transfers $q$, Eq.~(\ref{eq:q}),
for a given $q_t$.
Since the NPB rate,
Eqs. (8), (19), involves  
$q^{-4}$ (squared Coulomb potential), the decrease of $q$ 
causes a noticeable growth of $L$
with $t$ at low $t$ (Figs.~5 and 6). With
increasing $t$, the `thermal screening' itself becomes
more important and suppresses $L$
at higher $t$.  

Figures 5 and 6 also show variation of the Coulomb
logarithm with $Z$.                                 
The Coulomb logarithm increases with $Z$ which is evidently 
the effect of the ion screening ($q_s/p_{\rm F}
\sim Z^{-1/3}$): the smaller the ratio 
$q_s/p_{\rm F}$,
the larger the reaction rate. Similar effects
in the low-temperature case are well known
(e.g., Itoh \& Kohyama 1983).

%******************************************************************
%                                                       FIGURE 5.
\begin{figure}
\begin{center}
\leavevmode
\epsfxsize=8.0cm \epsfbox{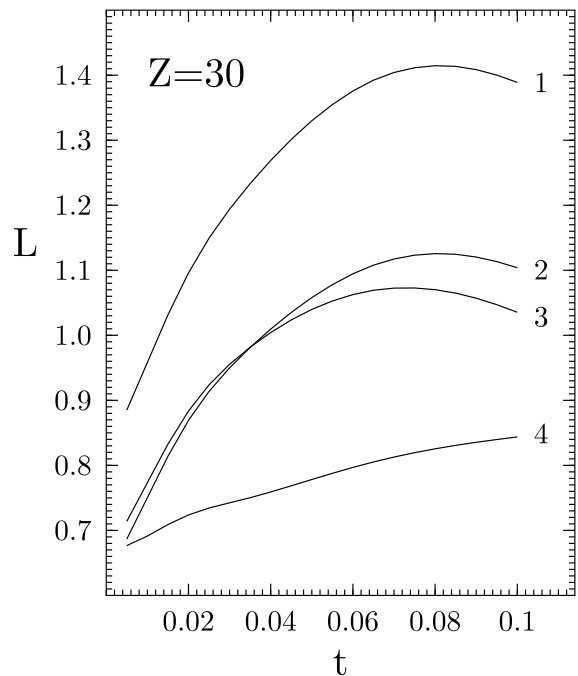}
\end{center}
\caption[ ]{
The NPB Coulomb logarithm $L$ vs. dimensionless temperature $t$ 
(\ref{eq:dimensionless}) for $Z=30$ nuclei. Curve 1
-- point-like nuclei ($\eta =r_c/a=0$);  2 --
finite-size nuclei with $\eta=0.2$; 3 --
same as 1 but in the Born approximation; 4
-- point-like nuclei,
non-Born corrections and thermal effects are neglected
(`standard approach' of previous works). 
}
\label{fig5}
\end{figure}
%******************************************************************

%******************************************************************
%                                                       FIGURE 6.
\begin{figure}
\begin{center}
\leavevmode
\epsfxsize=8.0cm \epsfbox{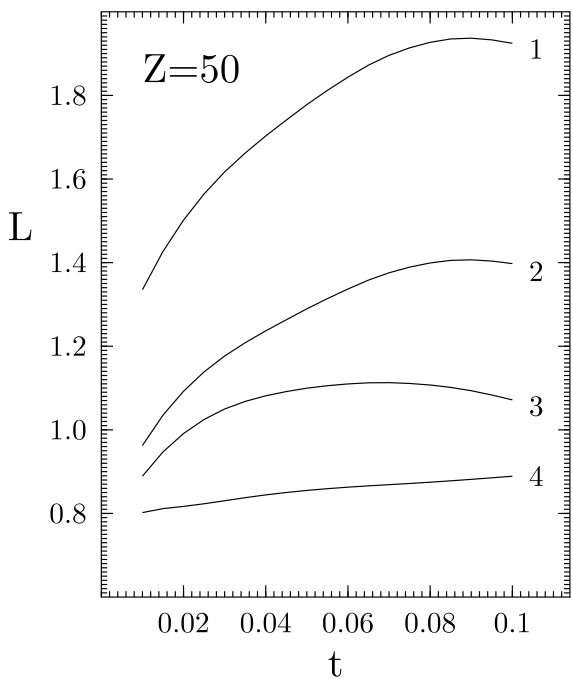}
\end{center}
\caption[ ]{
Same as in Fig.~5, but for $Z$=50.
}
\label{fig6}
\end{figure}
%******************************************************************

The calculations show that finite size of nuclei
(nuclear formfactor) becomes
significant for $\eta \ga (0.1 - 0.2)$. With increasing
$\eta$, the Coulomb logarithm noticeably decreases
since the nuclear formfactor introduces an
effective screening of the Coulomb interaction
(Sect.~3.2). Finite-size effects are negligible
in the outer crust of a neutron star, where $\eta \ll 1$,
but they are strong in the inner crust, where the
atomic nuclei occupy a substantial fraction of volume.

%%%%%%%%%%%%%%%%%%%%%%%%%%%%%%%%%%%%%%
Finally, let us emphasize the importance of the 
non-Born corrections. 
According to Doggett \& Spencer (1956),
the non-Born correction factor $R_{\rm NB}$
increases the electron - nucleus scattering cross section,
$R_{\rm NB}>1$, and the increase is larger for higher $Z$.
Accordingly, an inclusion of
this factor amplifies quite noticeably the Coulomb logarithm
as seen from Figs.~5 and 6. For $Z \ga 30$, this effect
is even more important than the thermal effect described above.

Comparing our improved values of $L$ which include the thermal
and non-Born corrections (curves 1 in Figs. 5 and 6)
with the non-corrected values (curves 4),
relevant to the ``traditional'' results,
we conclude that our corrections amplify $L$ and
the NPB rate, typically, by a factor of 1.5 -- 2.5.

%   New fit:
%   Liquid:=(0.269+20*t+0.0168*Z+0.00121*f
%     - 0.0356*Z*f+0.0137*Z*Z*t+1.54*Z*t*f)/
%    Pow((1+180*t*t+0.483*t*Z+20*t*Z*f*f+4.31E-0005*Z*Z),0.75);
%   {AvErr=0.0085  MaxErr=0.0233 at Z=10 t=0.001 f=0.2 Total N = 105}
For a practical use, we can propose an analytic fit:
\begin{eqnarray}
       L & =  & A/B^{3/4},
\label{eq:Fit} \\
       A & = & 0.269+ 20 t+ 0.0168 Z+0.00121 \eta
     - 0.0356 Z \eta
\nonumber \\
     & + & 0.0137 Z^2 t + 1.54 Z t \eta,
\nonumber \\
       B & = & 1+180 t^2 +0.483 t Z + 20 t Z \eta^2+
               4.31 \times 10^{-5} Z^2.
\nonumber
\end{eqnarray}
This formula reproduces all calculated values of $L$ ($Z \leq 50$,
$ t \la 0.1$,  $\eta \la 0.2$)
with the mean error of about
1\%, and with the maximum error of 2.3\% at $Z=10$, $t=0.001$
and $\eta = 0.2$.
%%%%%%%%%%%%%%%%%%%%%%%%%%%%%%%%%%%%

Thus the NPB energy loss rate $Q_{\rm Brem}$ can be easily
calculated from Eqs.~(\ref{eq:QthroughL}) and (\ref{eq:Fit})
for any model of dense matter (provided the values of $\rho$,
$T$, $n_e$, $Z$, $A$, $X_A$, and $r_0$ are specified).
If matter consists of nuclei of various species $(A,Z)$,
one should sum over the species in
Eq.~(\ref{eq:QthroughL}). Actually,
NPB from a multi-component mixture
of nuclei deserves a separate study. We expect, however,
that our results (obtained for one component plasma of nuclei)
can be used, at least semi-quantitatively,
for the mixtures as well.

Note that Cazzola et al. (1971)
and Munakata et al. (1987) considered the
NPB of non-degenerate and weakly
degenerate electrons.
In this case the thermal shell
washes out the Fermi surface, and
the thermal effects are naturally
implanted in the equations.
However the above authors analyzed ideal plasma
of ions $\Gamma \ll 1$ (neglecting the ion screening)
and weak degeneracy ($T/T_{\rm F} \ga 0.3$) -- the conditions
which are not relevant for applications and
opposite to those studied in our work.

Figure 7 shows the density dependence of the NPB energy loss rate
$E_{\rm Brem}=Q_{\rm Brem}/\rho$ 
% (ergs~s$^{-1}$~g$^{-1}$) 
(logarithmic scale)
for three models of dense matter (Sect.~2)
at several $T$.
For the ground state and accreted matter,
the NPB is produced by electron scattering on
nuclei $(A,Z)$ of one species (Sect.~2).
In the case of the hot matter, we take into account the
contribution of nuclei, protons and $\alpha$ particles.
Note that self-consistent models
of accreted matter correspond to
$T \ga 10^8$~K (e.g., Miralda-Escud$\acute{e}$ et al. 1990).
We use such a model at higher $T$
for illustrative purpose, to show how possible scatter
in nuclear composition affects the NPB rate.
The rates for accreted and ground state matters
are shown at $T \leq 5 \times 10^9$~K, while the rate
in the hot matter is displayed at $T \geq 5 \times 10^9$~K
(Sect.~2).
If $T \sim 5 \times 10^9$~K, all three models are possibly
not very accurate. We present three curves
at $T = 5 \times 10^9$~K
to visualize the effects of nuclear composition.
The NPB rates at this temperature are shown separately
in Fig.~8 in natural scale. 
It is seen that they differ by a factor of 2 -- 4.

At $T=10^9$~K and $2 \times 10^9$~K,                        % ??? 
in the considered density region 
$10^8 - 10^{13}$~g~cm~$^{-3}$
the matter may be in liquid and crystal phases.
The quantity $E_{\rm Brem}$ in the crystal phase was calculated
using simple approximations
obtained recently by two of the authors
(Yakovlev \& Kaminker 1996)                 % ???
for neutrino pair emission due to the
electron-phonon scattering.
The ground state matter is crystallized for 
$\rho > 2.35 \times 10^{10}$~g~cm~$^{-3}$ at  $T=10^9$~K
(not for all densities, see below)
and for  $\rho > 1.3 \times 10^{11}$~g~cm~$^{-3}$  
at  $T=2 \times 10^9$~K.
The accreted matter is crystallized for $\rho > 2.14 
\times 10^{11}$~g~cm~$^{-3}$  (not for all densities)
only at $T=10^9$~K.  
The melting temperature for the accreted matter
is smaller (Fig.~2) due to lower values of $Z$,
and the crystallization occurs at higher $\rho$
(if $T$ is fixed). It should be noticed that
there are two types of jumps of the NPB rate
for the ground state and accreted models.
The jumps of the first type 
result  from the jumps of $Z$ and $A$ (Sect.~2, Figs.~1, 2).
Such jumps take place at $T=10^9$~K, $2 \times 10^9$~K
and $5 \times 10^9$~K (Figs. ~7, 8).
The other more pronounced jumps come from
transitions from liquid phase to crystal and back
at some densities. One may  see single `back' transition 
from crystal to liquid for the ground state matter  
and a few ones for the accreted matter 
at $T=10^9$~K (Fig.~7).
The `back' transitions come from nonmonotonic  
character of melting
temperature due to the jumps of $Z$ (see Fig.~2).
When $T$ is fixed  and
$\rho \ga 3 \times 10^9$ g~cm$^{-3}$,
the NPB rate for the
ground state matter in the liquid phase
is, generally, several times higher
than for the accreted matter
due to larger values of $Z$ (Sect.~2). 

%%%%%%%%%%%%%%%%%%%%%%%%%%%%%%%%%%%%%%%%
Note, however, the striking general similarity of the NPB rates
in the liquid and solid phases (Fig.~7).
It results from the similarity of the expressions
for the NPB rates (cf. Eq. (21) with  Eqs. (13) and (14) of
Yakovlev \& Kaminker 1996) and reflects
common properties of strongly coupled Coulomb
liquid and high-temperature Coulomb crystal.
%%%%%%%%%%%%%%%%%%%%%%%%%%%%%%%%%%%%%%%%%%

The NPB rate in
the hot matter for $T=1.2 \times 10^{10}$~K (Fig.~7)
is broken at densities $\rho \la 2 \times 10^8$ g~cm$^{-3}$
at which the electron degeneracy becomes low and our results
are invalid. Note a sharp drop of the NPB rate
at $\rho \la 10^9$ g~cm$^{-3}$ for this $T$.
It occurs due to dissociation of nuclei (Fig.~3): the contribution
of nuclei into the NPB becomes negligible but a smaller contribution
of protons and $\alpha$-particles is available.
The drop clearly indicates the importance of coherence
effect ($Q_{\rm Brem} \propto Z^2$) which amplifies significantly
the NBP rate in the presence of high-$Z$ nuclei.
For other $T$ and $\rho$, the contribution of protons and
$\alpha$-particles is insignificant.

If $Z$, $A$, $X_A$, and $L$ were independent of density,
the NPB rate $E_{\rm Brem}=Q_{\rm Brem}/ \rho$ would also be density
independent (at fixed $T$). However, as seen from Figs.~7 and 8,
the NPB energy loss rate $E_{\rm Brem}$ decreases mainly with $\rho$ at
$\rho \ga 10^{10}$~g~cm$^{-3}$. This decrease
is explained mostly by the lowering of  $Z^2/A$,
(Figs.~1 and 4),  fraction of nuclei 
$X_A$, and of the Coulomb logarithm $L$.

%******************************************************************
%                                                       FIGURE 7.
\begin{figure}
\begin{center}
\leavevmode
\epsfxsize=8.0cm \epsfbox{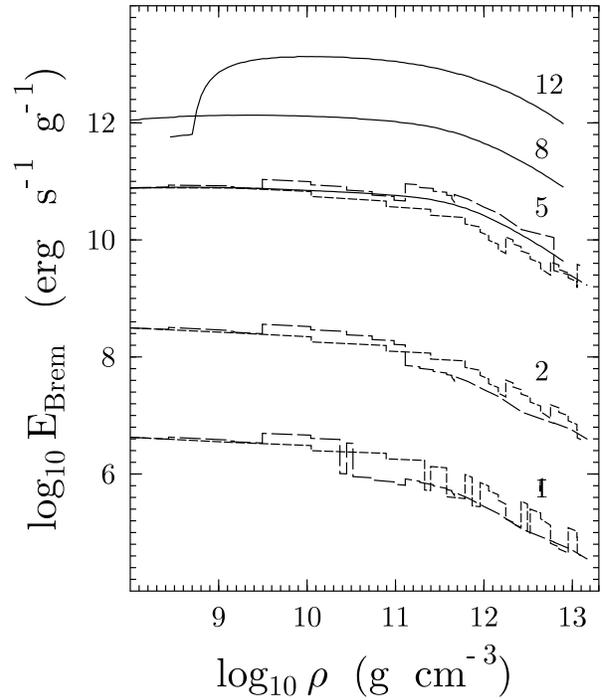}
\end{center}
\caption[ ]{
The NPB energy loss rate $E_{\rm Brem}=Q_{\rm Brem}
/\rho$ (in ${\rm erg~g^{-1}~s^{-1}}$, logarithmic scale)
vs. density (in g~cm$^{-3}$) for
$T_9$=1, 2, 5, 8, 12 (figures near curves).
Solid line --
hot matter; long dashes -- ground state matter,
short dashes -- accreted matter.
}
\label{fig7}
\end{figure}
%******************************************************************

%******************************************************************
%                                                       FIGURE 8.
\begin{figure}
\begin{center}
\leavevmode
\epsfxsize=8.0cm \epsfbox{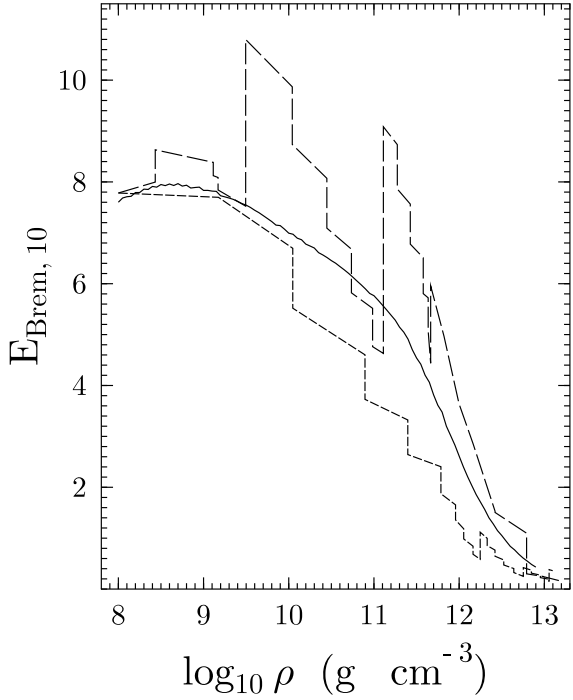}
\end{center}
\caption[ ]{
Same as in Fig.~7, but 
for $T=5 \times 10^9$~K (linear scale). The plotted 
quantity is
$E_{\rm Brem,10}$  (the loss rate $E_{\rm Brem}$ in units of
$10^{10}$~ergs~s$^{-1}$~g$^{-1}$).
}
\label{fig8}
\end{figure}
%******************************************************************

% Section 4 *************************************************
\section{Conclusions}
We have analyzed nuclear composition of neutron star
crusts (Sect. 2) for three models of dense matter:
cold catalyzed matter, accreted matter, and hot matter
in thermodynamic equilibrium. The first and second models are
valid for $T \la 5 \times 10^9$~K, while the third one
is appropriate for higher temperatures. Nuclear composition ($A,Z$),
melting temperature and other
properties of dense matter are different
for these models.

In Sect.~3 we have reconsidered the neutrino-pair
bremsstrahlung (\ref{eq:Reaction}) of relativistic
degenerate electrons due to scattering on atomic
nuclei at $T>T_{\rm m}$. We have shown that the neutrino
energy generation rate can be expressed (Eq.~(\ref{eq:QthroughL}))
through a Coulomb logarithm $L$, which varies slowly
in stellar matter. We have
obtained a general expression (\ref{eq:L}) for $L$, 
which takes into account two effects neglected in previous works.
First, it includes the non-Born corrections, and, second,
it describes the thermal effects associated with the finite width
of the shell around the electron Fermi surface.
The thermal effects are shown to
be quite important at moderate electron degeneracy,
and they lead to the appearance of the new `moderate
temperature regime' (Sect.~3.4) overlooked
in previous studies of NPB from relativistic
degenerate electrons. We have calculated
the Coulomb logarithm numerically for
possible parameters of dense matter in neutron
star crusts, and found a simple analytic
fit (\ref{eq:Fit}).

Note that NPB of electrons in a Coulomb liquid of
atomic nuclei is similar to NPB at $T<T_m$
produced due to Coulomb scattering of electrons on
`charged impurities' -- nuclei ($A_{\rm imp},Z_{\rm imp}$) immersed
accidentally in a lattice of bulk nuclei ($A,Z$).
When the nuclei crystallize ($T<T_{\rm m}$) but temperature
does not drop much below $T_{\rm m}$, the most important
is the neutrino pair generation due to electron phonon
scattering in solid matter (this follows from the recent
results of Pethick \& Thorsson 1994).  
The neutrino energy generation rate drops rapidly
with decreasing $T$ at $T \ll T_{\rm m}$ 
(e.g., Itoh et al. 1984b, Yakovlev \& Kaminker 1996),        % ???
and the Coulomb impurity scattering can be dominant
even for not very low $T$. 
The energy generation rate $Q_{\rm imp}$ is
obviously given by  the expression similar to 
(\ref{eq:QthroughL}) but with
$n_{\rm i}$ replaced by the impurity number density
$n_{\rm imp}$, and with $Z^2$ replaced by $(Z_{\rm imp}-Z)^2$.     
The expression for $Q_{\rm imp}$ contains a Coulomb logarithm
$L_{\rm imp}$ which should be quite similar to the
Coulomb logarithm (\ref{eq:LlowT}) 
in liquid matter (although the thermal effects
become unimportant).
The major difference is in the Coulomb screening.
In solid matter, the screening is described by the impurity
structure factor $S_{\rm imp}(q)$ which takes into account              
impurity distribution over lattice sites.
If the impurity correlation length is large,
the main contribution into the screening comes
from the electrons.

The results of the present article can be useful
for numerical modeling various phenomena related to
the thermal evolution of neutron stars. First of all,
we should mention cooling of young
neutron stars (of age $t \la$ (1 -- 10$^3$) yrs)
where internal thermal relaxation is
not achieved (Lattimer et al. 1994).
The thermal relaxation is accompanied by the 
propagation of the cooling  wave from the interior to
the surface. The associated variations of surface
temperature are, in principle, observable.
The dynamics of thermal relaxation is sensitive
to the properties of matter in the neutron star crust,
particularly to nuclear composition and neutrino
generation mechanisms. In addition, the above results
are useful for studying thermal evolution
of accreting neutron stars (see, e.g.,
Miralda-Escud$\acute{e}$ et al. 1990, and references
therein).

\acknowledgements
This work was supported partly by
INTAS, grant No. 94-3834,
the Russian Basic Research Foundation,
grant No. 96-02-16870a,
and by the ESO C\&EE Programme, grant No. A-01-068.
We acknowledge also the support of the Polish State Committee
for Scientific Research (KBN), grant No. 2 P304 014 07.

%   APPENDIX A
\appendix
\addtocounter{section}{1}
\section*{Appendix A: NPB matrix elements and energy loss rate}
\setcounter{equation}{0}
Using the notations introduced in Sect.~3.1,
the NPB energy loss rate in the Born approximation
can be generally written as ($\hbar=k_{\rm B} = c=1$):
\begin{eqnarray}
      Q_{\rm Brem} & = & \frac{n_i}{(2 \pi)^{11}}
            \int \! \dd {\bf p} \! \int \! \dd {\bf p}'
            \! \int \! \dd {\bf k}_{\nu} \! \int \! \dd {\bf k}'_{\nu}
                     \nonumber \\
                     & \times   &
            \delta(\ep - \ep' -\om) \om f(1-f') W ,
\label{eq:AQgeneral}
\end{eqnarray}
where 
\begin{equation}
      W= \frac{G_F^2}{2} \; \frac{1}{(2 \omega_{\nu})(2 \omega'_{\nu})
         (2 \ep)(2 \ep')} \; \sum_{\sigma, \nu} | M |^2,
\label{eq:AW}
\end{equation}
$|M|^2$ is the squared matrix element. Summation is over the
electron spin states $\sigma$ before and after scattering
and over neutrino flavors ($\nu_e$, $\nu_\mu$, $\nu_\tau$).
When the neutrino energies are much lower
than the intermediate boson mass
($\sim$80 GeV) the standard approach yields
\begin{eqnarray}
     \sum_{\sigma} |M|^2 & = &
              \int \dd {\bf q} \, |U({\bf q})|^2 \,
              \delta({\bf p} - {\bf q} - {\bf p}' - {\bf k})
\nonumber \\
     & \times &  {\rm Tr}(\hat{K_1} O^\alpha \hat{K_2} O^\beta)
                        \nonumber   \\
                      & \times &
              {\rm Tr} \left[ (\hat{P}'+m) L_\alpha
              (\hat{P}+m) \bar{L}_\beta \right] ,
\label{eq:AM} \\
    O^\alpha & = & \gamma^{\alpha} (1+ \gamma^5),   % \hspace{5mm}
                      \nonumber \\
                       &    &
              L_{\alpha} = \Gamma_{\alpha} G(P-Q) \gamma^0 +
              \gamma^0 G(P'+Q) \Gamma_{\alpha},
\label{eq:AO} \\
    G(P) & = & \frac{\hat{P} + m}{P^2 - m^2},
              \hspace{5mm} \Gamma^{\alpha} = C_V \gamma^{\alpha} +
              C_A \gamma^{\alpha} \gamma^5.
\label{eq:AG}
\end{eqnarray}
Here  $U({\bf q})$ is the Fourier transform
of the Coulomb potential (Sect.~3.2),
$m$ is the electron mass, $G(P)$
is the free-electron propagator, $\gamma^\alpha$ is a Dirac 
 matrix, upper bar denotes Dirac conjugate, and
$\hat{P} = P_\alpha \gamma^\alpha$
(Berestetskii et al. 1982).
 Furthermore, $C_V$ and $C_A$ are, respectively,
the vector and axial vector
weak interaction constants. For the
emission of electron neutrino
(charged + neutral currents), one has
$C_V = 2 \sin^2 \theta_W +0.5$ and $C_A= 0.5$,
while for the emission of muonic or tauonic
neutrinos (neutral currents only),
$C'_V = 2 \sin^2 \theta_W - 0.5$ and $C'_A = -0.5$.
Here $\theta_W$ is the Weinberg angle,
$\sin^2 \theta_W \simeq 0.23$.

Using the well known identity
\begin{eqnarray}
                    &      &
           \int \! \dd {\bf k}_\nu \int \! \dd {\bf k}'_\nu \,
           \delta^{(4)}(K - K_1 - K_2) \,
           \frac{ K_1^\alpha {K}_2^\beta}{\omega_\nu \omega'_\nu}
             \nonumber \\
                    &  =  &
          \frac{\pi}{ 6} (K^2 g^{\alpha \beta} + 2 K^\alpha K^\beta),
\label{eq:AIdentity}
\end{eqnarray}
we obtain
\begin{eqnarray}
        Q_{\rm Brem} & = & \frac{G_F^2 \,n_i }{12 (2 \pi)^{10}}
             \int \! \dd {\bf k} \int \! \dd {\bf p}
             \int \! \dd {\bf p}' \, |U({\bf q})|^2
                               \nonumber  \\
                         & \times    &
             f (1-f') \frac{\om}{ \ep \ep'} J \, ,
\label{eq:AQ1} \\
          J  & = & \sum_\nu (K^\alpha K^\beta - K^2 g^{\alpha \beta})
                               \nonumber  \\
                         & \times   &
                 {\rm Tr} \left[(\hat{P}' + m) L_\alpha (\hat{P} + m)
                 \bar{L}_\beta \right] .
\label{eq:AJgeneral}
\end{eqnarray}
The integration in (\ref{eq:AQ1}) is to be
carried out over the domain where $K^2 \geq 0$;
$g^{\alpha \beta}$ is the metric tensor.

Let us introduce the notations
\begin{eqnarray}
        u &= & (P' + K)^2 - m^2 = 2 P' K + K^2,       % \hspace{5mm}
              \nonumber  \\
        v & = & -(P - K)^2 + m^2 =   2 P K - K^2.
\label{eq:Auv}
\end{eqnarray}
Tedious but straightforward calculations yield:
\begin{eqnarray}
      J & = &  C_+^2 J_+ + C_-^2 J_-,            % \hspace{5mm}
            \nonumber \\
        C_+^2 &  = &  \sum_\nu (C_V^2+ C_A^2),         %  \hspace{5mm}
             \nonumber \\
        C_-^2 &  = &  \sum_\nu (C_V^2- C_A^2),
\label{eq:AJ1}
\end{eqnarray}
where
\begin{eqnarray}
    J_+ & = & 8 m^2 - 4(m^2+K^2)
               \left(\frac{u}{v} + \frac{v}{u} \right)
     \nonumber \\
     & + &  \left( \frac{4}{v} - \frac{4}{u} \right)
            \left[ K^2 \left( 4 PP' - 8 \ep \ep'+ m^2
                      \right.   \right.
                      \nonumber  \\
                      &  +   &
                     \left. \left.
         K^2 -   4 \om (\ep - \ep') \right)  - 4m^2 \om^2 \right]
                       \nonumber \\
     & + & \frac{4}{u^2} K^2 (m^2 - K^2)
           (4 \ep \ep' - 2 PP'
                      \nonumber \\
                      &  +    &
                 2 m^2 + 4 \ep \om - K^2 )
                        \nonumber \\
     & + & \frac{4}{v^2} K^2 (m^2 - K^2)
           (4 \ep \ep' - 2 PP'
                        \nonumber \\
                     &   +   &
         2m^2 - 4 \ep' \om  - K^2 )
     \nonumber \\
     & + & 4 K^2 (m^2 -K^2) \left(
              \frac{u}{v^2} - \frac{v}{u^2} \right)
             \nonumber \\
     & + & \frac{8}{uv} \left[ 4 K^2 (PP'- m^2)(2\ep \ep' -PP'-m^2)
                       \right.
                       \nonumber   \\
                          &   -   &
               2 m^2 K^2 (2\ep \ep' - PP' + m^2)
     \nonumber \\
     & + & 4 \om^2 (m^2 + K^2) PP' - 2 m^2 \om^2 (2 m^2 + K^2)
     \nonumber \\
     & + & \left. K^4 (3 \ep^2 + 3 \ep'^2 +2 \ep \ep' -2PP' -
                    m^2 - \om^2)    \right]
\label{eq:AJplus}
\end{eqnarray}
and
\begin{eqnarray}
      J_- & = & - 8 m^2 + 4 m^2 \left(\frac{u}{v} + \frac{v}{u}\right)
                   \nonumber \\
                       &   +   &
              \left( \frac{4}{v} - \frac{4}{u} \right)(4 \om^2 + K^2)
                    \nonumber     \\
      & - &  \frac{12}{u^2} m^2 K^2 (4 \ep \ep' - 2PP'
                          \nonumber \\
                       &   +    &  2m^2 +
                        4 \ep \om - K^2)
                          \nonumber     \\
      & - &  \frac{12}{v^2} m^2 K^2 (4 \ep \ep' - 2PP'
                              \nonumber \\
                             &  +    &  2m^2 -
                        4 \ep'\om - K^2)
      \nonumber      \\
      & - &   12 m^2 K^2 \left( \frac{u}{v^2} - \frac{v}{u^2} \right)
      \nonumber       \\
      & + &   \frac{8 m^2}{uv} \left[ K^2 (12 \ep \ep' + 2 PP' -2m^2
               + K^2)    \right.
                  \nonumber \\
             &  -   &  \left.  2 \om^2 (2PP' - 2 m^2 + K^2) \right] \,.
\label{eq:AJminus}
\end{eqnarray}
%

%%%%%%%%%%%%%%%%%%%%%%%%%%%%%%%%%%%%%%%%%%%%%%%%%%%%%%%
Equations (\ref{eq:AQ1}), (\ref{eq:AJgeneral})
(\ref{eq:AJplus}) and (\ref{eq:AJminus})
determine
the NPB rate for any degree of electron degeneracy
and relativism. The equations yield the
solution of the problem in a compact
and convenient form which allows one to consider
easily different limiting cases.
In a hot, non-degenerate plasma, where
a significant number of
positrons can be present in addition to the electrons,
the contribution from positrons should be added to $Q_{\rm Brem}$
in the straightforward manner.
%%%%%%%%%%%%%%%%%%%%%%%%%%%%%%%%%%%%%%%%%%%%%%%%%%%%%%

Analyzing Eqs.~(\ref{eq:AJplus}) and (\ref{eq:AJminus})
one can show that the main contribution into NPB
of relativistic and degenerate electrons comes from $J_+$.
Accordingly, we keep only this term in
Eq.~(\ref{eq:QthroughL}).

In our case, the neutrino-pair momentum is
thermal, $k \la T \ll p_{\rm F}$ but the elastic
electron - nucleus momentum transfer can be much larger,
$q_t \sim p_{\rm F}$. First consider the case
when $q \gg T$ or $q \gg k$.
Then we can put ${\bf q} \approx {\bf p} - {\bf p}'$,
and the Coulomb scattering
is, to a very good approximation, elastic.
A careful analysis of various terms in (\ref{eq:AJplus})
shows that the main contribution comes from the term
\begin{eqnarray}
         J_+ &  = &  \frac{8 K^2 (P P')[2 \ep \ep' - (P P')]}
                  {(P'K) (PK)}
                      \nonumber \\
                      &  = &
         \frac{8 ( \om^2 - {\bf k}^2)[(\ep \ep')^2 - ({\bf p}{\bf p}')^2]}
         {(\ep' \om - {\bf p}'{\bf k})(\ep \om - {\bf p}{\bf k})}.
\label{eq:AJlargeq}
\end{eqnarray}
Here we have adopted the natural approximation: $u=2P'K$, $v=2PK$.
In addition, we can set $\ep = \ep' = \mu$,
$p=p'=p_{\rm F}$  and  $({\bf q}{\bf p})/p = q^2/2p$ (see Sect.~3.3),
in smooth functions of electron energy
and momentum. One can easily show that
$(\ep \ep')^2 - ({\bf p}{\bf p}')^2 \approx  p^2 \, q^2 \,
[1 - q^2/(2 p^2)] $, $u \approx 2p q_r$ and 
$v \approx 2p(q_r - ({\bf q}_t {\bf k})/p)$, where we use the
notations of Sect.~3.3. Finally, 
for the case of $T \ll q $ we obtain
\begin{equation}
      J_+= 8 {q_t^2 \over q_r}  (\omega^2 -k^2)
        \left(1 - {q_t^2 \over 4p^2} \right)
         \left(q_r- {{\bf k} {\bf q}_t \over p} \right)^{-1}.
\label{eq:AJ}
\end{equation}

Now consider $J_+$ in the approximation of small-angle
scattering ($q \ll p_{\rm F}$).
Then $k$ can be comparable to $q$, and
more terms should be kept in $J_+$. An analysis of
Eq.~(\ref{eq:AJplus}) shows that we must retain the terms
\begin{eqnarray}
      J_+ & = & \frac{32 \, K^2}{uv} [(\ep \ep')^2 - ({\bf p} {\bf p}')^2]
           \, - \, 4 K^2
            \left( \frac{v}{u} + \frac{u}{v} \right)
                     \nonumber \\
                     &  + &
          16 K^2 \left( \frac{1}{u} - \frac{1}{v} \right)
           ( \ep \ep' +  {\bf p} {\bf p}')
                     \nonumber \\
         &   -   &  8 \, K^4 ( \ep \ep' + {\bf p} {\bf p}')
         \left( \frac{1}{u^2} + \frac{1}{v^2} \right)
                        \nonumber \\
                        &  +  &
         \frac{16 K^4}{uv} (\ep + \ep')^2 \,.
\label{eq:AJsmallqInter}
\end{eqnarray}

Using energy-momentum conservation
and the inequalities  $k \ll p$ and $q \ll p$
and introducing the notations of Sect.~3.3, we obtain
\begin{eqnarray}
        \om &\approx & q_r + k_r,
     \nonumber \\
        u & \approx & v  \approx 2p q_r,
     \nonumber \\
       K^2 & = & \om^2 - k_r^2 - k_t^2 \approx
        k_0^2 - k_t^2,
     \nonumber \\
         J_+ & = &  8 (\omega^2-k^2) \frac{q_t^2}{q_r^2}.
\label{eq:AJsmallq}
\end{eqnarray}

Let us compare Eqs.~(\ref{eq:AJ}) and (\ref{eq:AJsmallq})
for $J_+$ obtained, respectively, in the domains of large and small $q$.
Since both domains overlap, we can describe
$J_+$ accurately for all $q$. However, we see that
Eq.~(\ref{eq:AJ}) coincides formally with (\ref{eq:AJsmallq})
in the domain of small $q$, where the terms $q_t^2/(4p^2)$
and ${\bf k} {\bf q}_t /p$  are negligible second-order corrections.
Thus we use Eq.~(\ref{eq:AJ})
for all $q$ in Sect.~3.3.

% APPENDIX B
\addtocounter{section}{1}
\section*{Appendix B. Thermal function $R_T(y,u)$}
\setcounter{equation}{0}
According to Sect.~3.3, the effects of inelasticity
in the electron-nucleus scattering are described by the
thermal function $R_T(y,u)$ given by Eq.~(\ref{eq:RT}).

Integrating over $u$ we obtain
\begin{eqnarray}
       \int_0^\infty \dd  u\; R_T(y,u)
       &  =   &  1 + \frac{2 y^2}{1 -  y^2} \ln y
                \nonumber    \\
             &  \rightarrow &
              \left\{ \begin{array}{cl}
                       1 & y \ll 1,  \\
                        \frac{1}{2}(1-y^2)  &  y \rightarrow 1.
               \end{array}
           \right.
\label{eq:BIntR}
\end{eqnarray}
In the limits of small and large $u$ we have
\begin{equation}
      R_T(y,u) = \frac{63}{160 \pi^2}\,u \frac{1-y^2}{y^2}, \; \;
      {\rm for} \; \; y^2 \gg u,
\label{eq:BAsym1}
\end{equation}
\begin{equation}
     R_T(y,u) = \frac{189}{32 \pi^6}\, u^2 {\rm e}^{-u/2}, \; \;
     {\rm for} \; \;  u \, (1-  y^2) \gg 2 \, y^2 .
\label{eq:BAsym2}
\end{equation}
If $y \ll 1$, Eq.~(\ref{eq:RT}) yields
\begin{equation}
      R_T(y,u) \rightarrow  R_T(u) =
      \frac{189}{16 \pi^6} \int_{u/2}^{\infty}
      \dd v \; \frac{v^2 \; }{{\rm e}^v -1}
      \left(v- \frac{u}{2} \right)^2.
\label{eq:BRTy0}
\end{equation}
The asymptotic forms are:
\begin{eqnarray}
      R_T(u)  =  \frac{189}{16 \pi^6} \Gamma(5) \zeta(5) & &
                  1 \gg u \gg y^2 ,
                  \nonumber  \\
          R_T(u)  =  \frac{189}{32 \pi^6}\;  
           u^2 \;{\rm e}^{- u/2} & &
                 u \gg 1,
\label{eq:BRTy0Asym}
\end{eqnarray}                                             
where $\zeta(x)$ is the zeta function.
Finally, for $(1 - y) \ll 1$  we have
\begin{eqnarray}
R_T(y,u) & \approx  &  \frac{189}{16 \pi^6} \; (1-y) \; u
                    \nonumber \\
                     & \times   &
           \int_{u/2}^{\infty}
           \dd v \; \frac{v^2}{{\rm e}^v - 1} \; 
             (v - \frac{u}{2}).
\label{eq:BRTy1}
\end{eqnarray}

We have calculated $R_T(y,u)$ numerically for wide ranges
of $y$ and $u$ (10 points of $y$ from $y$=0 to 0.9;
10 points of $u$ from $u$=0.1 to 30).
We have found the following fitting expression
\begin{eqnarray}
       R_T(y,u)& = & u (1-y^2)\; {1+0.5186 y^2 \over u+ 2.608 y^2} \;
       { F \over G},
\nonumber \\
       F & = & 0.3058 + 4.331 y^2 + F_1 u \, {\rm e}^{u/2},
\nonumber \\
       G & = & 1+ 58.05 y^2 + G_1 u \, {\rm e}^u,
\nonumber \\
       F_1& = &2.949 - 2.963y^2 + u F_2,
\nonumber \\
       F_2& = &0.7184 -0.8565y^2+0.06019u
                \nonumber \\
                  &  +  &
           0.07671y^2u -0.0007771 u^2y,
\nonumber \\
       G_1& = &9.797-6.502y^2.
\label{eq:BfitR}
\end{eqnarray}
The mean error of the fit is 3.6\%, and the maximum error
of 8.1\% takes place at $y$=0.5 and $u$=4.
This expression has been used in calculations presented in
Sect.~3.5.

%  APPENDIX C
\addtocounter{section}{1}
\section*{Appendix C: Dynamic screening}
\setcounter{equation}{0}
In principle, the ion screening of the Coulomb interaction can be
dynamic. For a strongly
coupled plasma of ions ($\Gamma \ga 1$), the dynamic screening can
be introduced through the dynamic structure
factor $S(q,\Omega)$ which satisfies the relationship
\begin{equation}
          S(q) = \int_{-\infty}^{+\infty}
          \frac{\dd \Omega}{2 \pi} \, S(q,\Omega).
\label{eq:CS}
\end{equation}
Here, $\hbar \Omega$ is the energy transferred
from electrons to nuclei.
Typical frequencies $\Omega$ of the dynamic response are
$\Omega \sim \om_p$, where $\om_p = \sqrt{4 \pi Z^2 e^2 n_i/ m_i}$
is the ion plasma frequency, and $m_i$ is the ion mass.
One can see that
$\hbar \om_p/ k_{\rm B} \la T_m \ll  T_{\rm F}$ in dense stellar
matter under consideration. 
In order to include the dynamic screening, we introduce $\Omega$
into the energy conserving
delta function in Eq.~(\ref{eq:AQgeneral}), and rewrite the static
structure factor $S(q)$ in the form (\ref{eq:CS}). 
Then the Coulomb logarithm for the relativistic degenerate electrons
reads (cf. Eq.~(\ref{eq:Lgeneral}))
\begin{eqnarray}
     L &  = &  \frac{189}{2^{11} \pi^9 T^6} \int_{-\infty}^{+\infty}
     \frac{\dd \Omega}{2\pi} \int \dd {\bf k} \; 
        \dd {\bf p} \; \dd {\bf p}'
        \frac{ S(q, \Omega) |F(q)|^2}{\ep \ep' q^4  | \epsilon(q) |^2 } 
                  \nonumber \\
                  & \times    &
                f(1-f') \; \om J_+.
\label{eq:CLgeneral}
\end{eqnarray}
This equation can be simplified as in Sect.~3.3.
Let us introduce the same quantities
${\bf q}= {\bf q}_t + {\bf q}_r $, 
${\bf k} = {\bf k}_t + {\bf k}_r$  and 
${\bf p}''= {\bf p} - {\bf q}_t = {\bf p}'+ {\bf k} + {\bf q}_r$ , 
where  $|{\bf p}''|= |{\bf p}|$.  
Using energy-momentum conservation and the inequalities
$ k \ll p$, $q_r  \ll p$ we obtain 
\begin{eqnarray}
        \Omega  +  \om & \approx & q_r + k_r  
        \nonumber    \\
         u & = & 2 P'K + K^2  \approx 2p(\om - k_r) = 2p(q_r - \Omega),
         \nonumber \\
         v & = & 2 PK - K^2 \approx 2p(q_r - \Omega - {\bf q}_t {\bf k}/ p)  ,
         \nonumber \\
        K^2 & = & \om^2 - k_r^2 - k_t^2 \approx
        k_0^2 - k_t^2.
\label{eq:ComuvK}
\end{eqnarray}
Here  $k_0^2= (q_r- \Omega)(2\om - q_r + \Omega)$, and
the condition  $K^2 > 0$  yields  $2\om > q_r - \Omega > 0$.
If $q \approx q_t$ and $q$  is much larger than $q_r$, $\Omega$ and $\om$,
we have (cf. Eq.~(\ref{eq:AJ}))
\begin{equation}
          J_+ \approx \frac{8 (k_0^2 - k_t^2) q_t^2}
          {(q_r - \Omega)(q_r - \Omega - {\bf q}_t {\bf k}/p)}
          \left( 1 - \frac{q_t^2}{4 p^2} \right).
\label{eq:CJ}
\end{equation}
On the other hand,
using the approximation of small-angle scattering
($q \ll p $)  and (\ref{eq:AJsmallq}),
we obtain (cf. Eq.~(\ref{eq:AJsmallqInter}))
\begin{equation}
           J_+ \approx  \frac{8 (k_0^2 - k_t^2)}{(q_r - \Omega)^2}
          \left[ q_t^2 + 2 \Omega (q_r - \Omega)  \right].
\label{eq:CJsmallq}
\end{equation}
Combining Eqs.~(\ref{eq:CJ}) and (\ref{eq:CJsmallq}), we can propose
the following interpolation 
\begin{eqnarray}
         J_+ & \approx & \frac{8 (k_0^2 - k_t^2)}
          {(q_r - \Omega)(q_r - \Omega - {\bf q}_t {\bf k} /p)}
          \left( 1 - \frac{q_t^2}{4 p^2} \right)
                    \nonumber \\
                    & \times   &
        \left[ q_t^2 + 2 \Omega (q_r - \Omega)  \right].
\label{eq:CJsingle}
\end{eqnarray}
Calculations similar to those in Sect.~3.3 allow us to express
$L$ in the form analogous to (\ref{eq:L}).
Using (\ref{eq:CJsingle}), we
come to a simple equation for the Coulomb logarithm:
\begin{eqnarray}
       L  & = & \int_{-\infty}^{+\infty} \frac{\dd \Omega}{2 \pi}
           \int_0^{2p_F} \; \dd q_t \; q_t \;
          \int_\Omega^\infty \frac{\dd q_r}{T}  \;
          [q_t^2+2 \Omega (q_r - \Omega)]
       \nonumber \\
         & \times & \frac{S(q, \Omega) |F(q)|^2 }{q^4 |\epsilon(q)|^2}
          R_T (y, u, w) R_{NB}(q_t).
\label{eq:CL}
\end{eqnarray}
Here  $q = \sqrt{q_t^2+q_r^2-q_t^2(q_r/p)}$,
and we introduce a new function
\begin{eqnarray}
       R_T(y,u,w)  & = & \frac{189}{16 \pi^6}
            \left\{  \int_{v_0}^{v_1} \dd v \;
            \frac{v(v + w)(v-v_0)^2 }{ {\rm e}^{v + w} -1}
                \right.
\nonumber \\
         & \times &  \left[1 - { v-v_0 \over 3 v_1 (1-y^2)} \right] 
\nonumber \\
       &+&   v_1(1-y^2) \,
            \int_{v_1}^\infty \dd v \;
            {v (v +w) \over {\rm e}^{v+w}-1 }
                     \nonumber \\
                     & \times     &  \left.
            \left[v - v_0 - {1 \over 3} v_1(1-y^2) \right] \right\},
\label{eq:CRT}
\end{eqnarray}
with $y=q_t/(2p_F)$,  $u=q_r/T$,  $w=\Omega/T$,  $v=\om/T$,
$v_0 =(u-w)/2$ and $v_1= v_0/y^2$.
This function describes the thermal effects (Sect.~3.3)
in the presence of the dynamic screening. 
Eq.~(\ref{eq:CRT}) yields
\begin{eqnarray}
  &  &  \int_0^{\infty} \dd u' R_T(y,w+u',w)   =
 \nonumber \\
  &  &  R_c(y) \, \frac{63}{8 \pi^6} \,
           \int_0^{\infty}  \dd v  \frac{v^4 ( v + w)}
    {{\rm e}^{v + w} - 1},
\label{eq:CIntR}
\end{eqnarray}
where $R_c(y)$ is defined by ~(\ref{eq:Rc}).
Consider the low-temperature case ($T \ll q_s$)
and take into account that $R_T(y,u,w)$ decreases
exponentially with increasing $(q_r - \Omega)$ at
$(q_r - \Omega) > T$  (see Sect.~3.4).
Then Eq.~(\ref{eq:CL}) reduces to
\begin{eqnarray}
   L & = & \frac{63}{8\pi^6} \, \int_0^{2p_F} \dd q_t \, q_t^3 \,
   \frac{|F(q)|^2  R_c(y) R_{\rm NB}(q_t)}{q^4 |\epsilon(q)|^2}
                 \nonumber \\
                 & \times   &
   \int_{-\infty}^{\infty} \frac{\dd \Omega}{2 \pi}  S(q, \Omega)
   \; \int_0^{\infty} \dd v \frac{v^4 (v + w)}{{\rm e}^{v+w} -1}.
\label{eq:CLlowT}
\end{eqnarray}
Let us assume further that the dynamic structure factor
$S(q,\Omega)$ decreases rapidly with the growth of $\Omega$
at low $\Omega$. Then
we can set $\Omega =0$ or $w=0$ everywhere in the
integrand of (\ref{eq:CL})  or (\ref{eq:CLlowT})
except in $S(q,\Omega)$. In this way we have
$R_T(y, u, w)=R_T(y, u)$  (cf. (\ref{eq:RT}) and (\ref{eq:CRT})).
The integration over $\Omega$ yields the static structure
factor $S(q)$ in accordance with ({\ref{eq:CS}}), and
the Coulomb logarithm ({\ref{eq:CL}}) transforms into the
static Coulomb logarithm ({\ref{eq:L}}).

Thus Eq.~(\ref{eq:CL}) extends the results of Sect.~3.3
to the case of the dynamic screening.
In the latter case the Coulomb logarithm is given by a
three dimensional integral which
could be computed if the dynamic structure factor were known.
However the detailed calculations of $S(q,\Omega)$ have not yet
been performed.
We expect that the formalism will be
useful in the future after $S(q,\Omega)$ is determined.
Our preliminary estimates indicate that the dynamic
effects are not very significant.

%************************************************************************

\end{document}